# Bias Control and Linearization of the Transfer Function of Electro-optic and Acousto-optic Modulators

Clemens Neumüller *, Frank Obernosterer **, Raimund Meyer **, Robert Koch *, Gerd Kilian *

* *Fraunhofer Institute for Integrated Circuits IIS, Erlangen, Germany*
  POC: robert.koch@iis.fraunhofer.de
** *Com-Research GmbH, Solutions for Communication Systems, Fürth, Germany*

## 1 Background

In several types of quantum computers light is one of the main tools to control both the position and the quantum state of the atoms used for computing. In practical systems laser light is applied to manipulate quantum states of qubits in the desired way. Beside physical effects like decoherence and quantum noise the precision of qubit manipulation has a significant impact on the achievable quantum computing error rate. Therefore, there are tight requirements on the laser regarding frequency precision, spectral bandwidth, output power stability etc.

One of the key optical components beside the laser is the optical modulator, which modulates or switches a constant power laser light in order to provide light pulses or pulse sequences with a desired envelope. Acousto-optic (AOM) and electro-optic (EOM) modulators can be applied, which are both voltage controlled. However, there is neither a simple linear relationship between their control signal and the precise modulator output, nor can they be considered to have time-invariant characteristics.

The aim of this document is to describe techniques to generate AOM and EOM control signals in such a way that almost arbitrary target output waveforms (i.e. optical power versus time) are achieved with high accuracy. This implicitly includes the capability of generating laser pulses with precisely defined energy.

## 2  Considered Subsystem

In this section, the considered control laser subsystem of a quantum computer is presented for the example of an acousto-optic modulator (AOM), which is restricted to the basic components required for the modulation of the laser light prior to its exposed to the target.

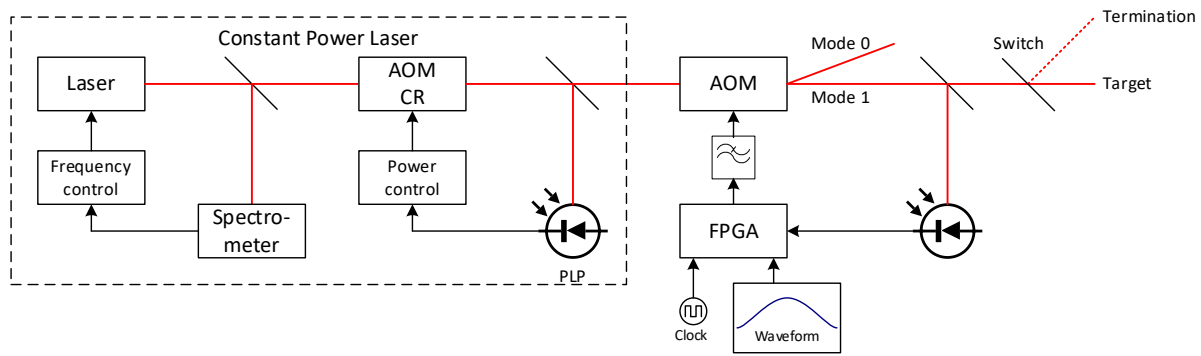

*Figure 1: Generation of modulated laser light using an acousto-optic modulator (AOM)*

The dashed line comprises the basic components of a constant power laser. The laser's internal frequency and power control loops are not considered within the scope of this work.

The constant power laser light is switched or modulated by an AOM or EOM (not depicted). In the following the more general term "modulation" will be used for the sake of simplicity, which also comprises "switching".

The modulation is controlled by a control voltage, which is generated by an FPGA device, including a D/A converter. The control voltage has to be generated in such a way that the optical power at the modulator output precisely corresponds to the desired shape, which may be a single pulse, a sequence of pulses or other arbitrary shapes. As will be shown later, the electro-optic transfer function is expected to have linear and nonlinear contributions, so that the envelope of the control voltage is different from the envelope of the modulator's output power. The process of generating a suitable input voltage for a desired output power shape is denoted as (digital) predistortion.
Due to temperature dependence and other effects, the behavior of the applied modulators is expected to be time-variant. Therefore, the predistortion parameters have to be adjusted at sufficiently short time intervals in order to permanently achieve the required accuracy of the output power shape, e. g. due to different environmental conditions. This is done by introducing a feedback loop used for parameter update. Figure 1 shows that a small fraction of the optical modulator output is directed to a photo diode measuring the actual power shape at modulator output. The voltage at the photo diode is fed back into the FPGA, where it is A/D converted and used for parameter update of the digital predistortion.

In total there are two basic algorithms involved:

- The digital predistortion algorithm, which converts the desired optical power shape into a corresponding modulator control voltage.
- The parameter update algorithm, which updates the parameters of the predistortion.

A more detailed description on this will be given in section 4.

# 3 Control Input to Optical Output Characteristics of investigated Modulators

This section describes the basic properties of electro- and acousto-optic modulators. The main focus is on the transfer function from control voltage input to optical output power, specifically on the expected linear and nonlinear distortion, time variance etc. A basic distortion model for EOM and AOM is developed, which allows for suitable compensation by appropriately adjusting the waveform of the control voltage (see section 4).

## 3.1 Electro-optic Modulator (EOM)

The fundamental phenomenon that accounts for Electro-Optic Modulators (EOM) or switches is the change in the refraction index $n(E)$ introduced by an applied electric field $E$. In the most general case, this effect is nonisotropic, and mainly contains linear (Pockels effect) components [1,2,5]. Thus, the linear electro-optic Pockels effect is widely used in optic applications and the medium is then known as Pockels medium [1] or cell. While many materials that can be used to produce low loss waveguides, most commonly lithium niobate (LiNbO$_3$, LN) is used.

A beam of light traversing a Pockels cell of length $l$ (see Figure 2) to which an electric field $E$ is applied undergoes a phase shift $\varphi = 2\pi \cdot n(E) \cdot l/\lambda_0$, where $\lambda_0$ is the free space wavelength. Using the refraction index, which is given in a Pockels medium [1] by

$$n(E) = n_0 - \Delta n(E) = n_0 - \frac{1}{2} n_0^3 \cdot r \cdot E, \qquad (1)$$

where $n_0 = n(E=0)$ describes the refraction index in the absence of an electric field $E$, $\Delta n$ characterizes the index change and the coefficient $r$ is the linear electro-optic coefficient (or Pockels coefficient). The phase shift $\varphi$ can then be described by

$$\varphi = \frac{2\pi \cdot n_0 \cdot l}{\lambda_0} - \pi \frac{n_0^3 \cdot r \cdot E \cdot l}{\lambda_0} = \varphi_0 - \Delta\varphi. \qquad (2)$$

If the electric field is obtained by applying a voltage $V(t)$ across two faces of the cell separated by the electrode distance $d$ (see Figure 2), then $E(t) = V(t)/d$, and Eq. (2) gives

$$\varphi = \varphi_0 - \Delta\varphi = \varphi_0 - \pi \frac{V(t)}{V_\pi}, \quad \text{where} \quad V_\pi = \frac{d \cdot \lambda_0}{l \cdot n_0^3 \cdot r} \quad \text{and} \quad \Delta\varphi = \pi \frac{V(t)}{V_\pi}. \qquad (3)$$

The parameter $V_\pi$, known as the half-wave voltage, is the voltage which shifts the phase by $\pi$. Eq. (3) expresses a linear relation between the optical phase shift $\varphi$ and the voltage $V(t)$. One can therefore modulate the phase of an optical wave by varying $V(t)$ that is applied across a material through which the light passes as shown in Figure 2. The parameter $V_\pi$ is an important characteristic of the modulator. It depends on material properties ($n_0$ and $r$), on the wavelength $\lambda_0$, and on the aspect ratio $d/l$.

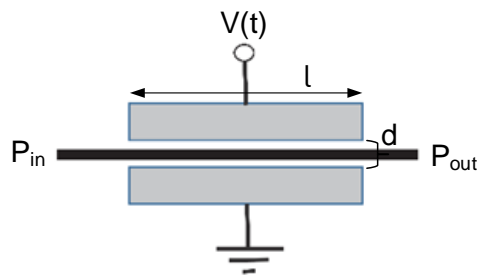

*Figure 2: An optical phase modulator based on the electro-optic effect [5]*

A phase delay by itself does not affect the intensity of a light beam. However, a phase modulator placed (in at least) one branch of an interferometer can act as an intensity modulator, as shown in Figure 3 for two variants of the Mach-Zehnder modulator (MZM). The first variant (Figure 3a) is the so-called Y-branch interferometer with one input and one output, in which the incoming optical wave is first divided into two arms by a Y-splitter and after the MZM recombined by a second flipped Y-splitter. The second variant (Figure 3b) shows the 3dB coupler MZM type with two inputs and two outputs. Light entering one of the inputs is split by a 3dB coupler equally into two partial waves, each of which passes through one of the interferometer arms. After this, the two waves are then recombined in another 50/50 3 dB coupler. The outgoing optical signals at the two outputs are complementary to each other. The main difference between both variants is the different number of outputs. If destructive interference of both partial waves takes place, then in the case of the Y-splitter MZM most of the light is emitted into the substrate, whereas in Figure 3b the light is emitted via the complementary output as a kind of waste port.

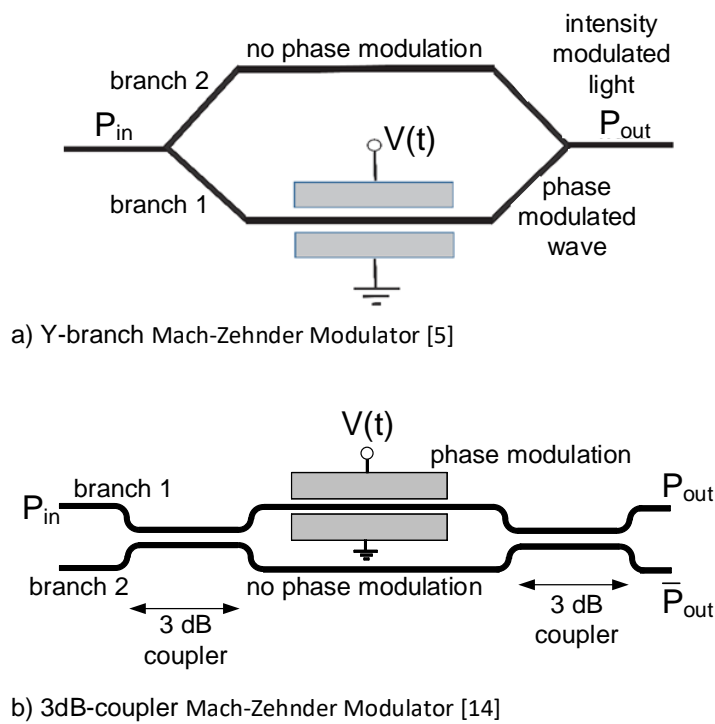

a) Y-branch Mach-Zehnder Modulator [5]

b) 3dB-coupler Mach-Zehnder Modulator [14]

*Figure 3: Different variants of an electro-optic (unbalanced) single-drive Mach-Zehnder modulator*

Since the phase modulator is only in one branch this variant is called (unbalanced) single-drive MZM [5]. If the optical beam splitter divides the optical input power equally $\left(P_1 = P_2 = \frac{1}{2}P_{in}\right)$, the output power $P_{out}$ is related to the input power $P_{in}$ by

$$P_{out} = \frac{1}{2}P_{in} + \frac{1}{2}P_{in} \cdot \cos\varphi = P_{in} \cdot \cos^2(\varphi/2) \qquad (4)$$

where $\varphi = \varphi_1 - \varphi_2$ is the total phase difference between the two branches. Optical losses at the device edges and during propagation are ignored in Eq. (4). Note, that in Eq. (4) the trigonometric half-angle identity $(1 + \cos x) = 2 \cdot \cos^2(x/2)$ has been used.

Because of the presence of the phase modulator in branch 1, according to Eq. (3) we have $\varphi_1 = \varphi_{1_0} - \pi\frac{V(t)}{V_\pi}$, so that $\varphi$ is controlled by the applied voltage $V$ in accordance with the linear relation $\varphi = \varphi_1 - \varphi_2 = \varphi_0 - \pi\frac{V(t)}{V_\pi}$, whereas the constant $\varphi_0 = \varphi_{1_0} - \varphi_2$ denotes the optical path difference. The transmittance $T(V)$ (or optical transfer function) of the interferometer is therefore a function of the applied voltage $V(t)$, i.e.

$$T(V) = \frac{P_{out}}{P_{in}} = \cos^2\left(\frac{\varphi_0}{2} - \frac{\pi}{2}\frac{V(t)}{V_\pi}\right) = \cos^2\left(\frac{\varphi_0}{2} - \frac{\Delta\varphi}{2}\right). \qquad (5)$$

This function is plotted in Figure 4 for an arbitrary value of $\varphi_0$. The device may be operated as a linear intensity modulator by adjusting the optical path difference so that $\varphi_0 = \pi/2$ (positions Quad±) and operating in the nearly linear region around T = 0.5. Alternatively, the optical path difference may be adjusted so that $\varphi_0$ is a multiple of $2\pi$. In this case $T(0) = 1$ and $T(V_\pi) = 0$, so that the modulator switches the light on and off as the voltage is switched between 0 and $V_\pi$.

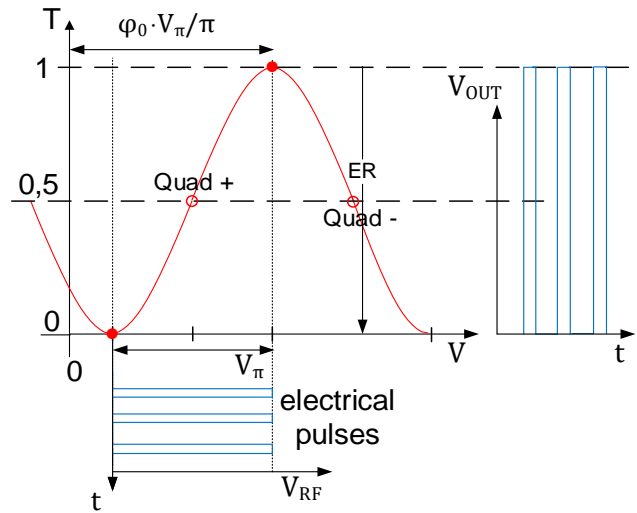

*Figure 4: Optical transmittance T versus applied voltage V for an arbitrary value of $\varphi_0$ and $P_1 = P_2$.*

The ideal EOM described in Eq. (4) with 50:50 optical splitter could provide full reflection (on-mode, $P_{out} = P_{max}$) and zero transmission (off-mode, $P_{out} = P_{min} = 0$). The ratio between $P_{max}$ and $P_{min}$ is called extinction ratio (ER) and in case of Eq. (4) resp. Eq. (5) ER is infinite. Even if a very high manufacturing accuracy is used in the production of MZMs, it cannot be balanced perfectly and thus only a finite extinction ratio can be achieved, which then leads to so-called leakage light. Lower ER result in higher Rayleigh backscattered noise and limits spatial resolution and sensing range.

For a finite extinction ratio ($P_{min} \neq 0$) Eq. (4) (together with Eq. (5)) changes to

$$P_{out}(t) = \frac{1}{2}P_{in} + \sqrt{P_1 \cdot P_2} \cdot \cos\left(\varphi_0 - \pi \frac{V(t)}{V_\pi}\right), \quad \text{for } (P_1 \neq P_2, \quad P_1 + P_2 = P_{in}). \tag{6}$$

The behavior of Eq. (6) is shown in Figure 5. The term $\sqrt{P_1 \cdot P_2}$ is the optical imbalance factor between the two branches.

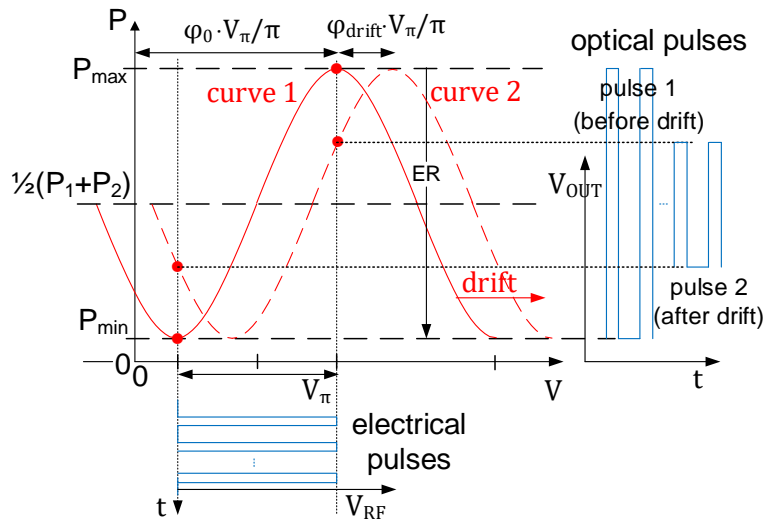

*Figure 5: Output power $P_{out}$ versus applied voltage V for an arbitrary value of $\varphi_0$ and $P_1 \neq P_2$.*

As mentioned above, a Lithium niobate Mach-Zehnder interferometer is never perfectly balanced. Mainly it is subject to a drift phenomenon, caused by a change of the effective refraction index $n$ of the waveguide's optical mode due to possible variations of the device's temperature (so called thermo-optic and pyroelectric effects), the electro-optic, photovoltaic and photoconductive properties of the material in the waveguide (so called photorefractive effects), aging, as well as static electrical charge accumulation. This drift $\varphi_{drift}$ causes the transfer function as shown in Figure 5 to move in the horizontal direction due to the introduced time dependence of the total phase difference in Eq. (6) to $\varphi(t) = \varphi_0 - \Delta\varphi(t) + \varphi_{drift}(t)$. The dashed red curve shows the behavior under the influence of a drift. In order to compensate this drift phenomenon in the MZM, one must apply the RF signal $V_{RF}(t)$ and a DC bias voltage $V_{DC}(t)$ to the modulator, as shown in Figure 6, in such a way that $V(t)$ in Eq. (6) equals

$$V(t) = V_{RF}(t) + V_{DC}(t). \tag{7}$$

Thus, the time-varying instantaneous total phase difference between both branches at time t can be decomposed as

$$\varphi(t) = \varphi_0 - \left(\varphi_{RF}(t) + \varphi_{DC}(t)\right) + \varphi_{drift}(t). \tag{8}$$

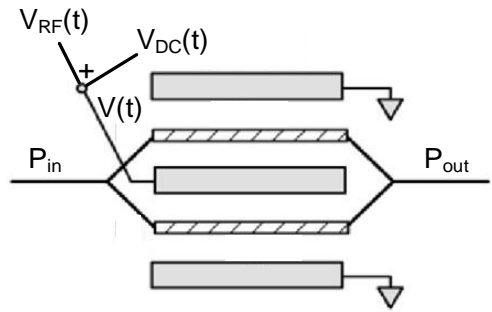

*Figure 6: An integrated-optical intensity dual-drive Mach-Zehnder modulator [15]*

Due to the drift, the modulator DC bias $\varphi_{DC}$ must be controlled to reach a stable operating point. This is usually done by specialized automatic feedback systems. There have been developed a lot of techniques to control the bias voltage of a MZM, see e.g. [3,4,5].

Figure 6 shows an example of a commercial Lithium niobate MZM in a so called dual-drive configuration, where now one sets of electrodes is used in each branch. Furthermore, the electrode configuration in both arms of the interferometer is such that the upper and lower arm undergo an equal but opposite phase delay change of ±π/2 (so called push-pull mode)

$$\Delta\varphi_1 = -\Delta\varphi_2 = \frac{\pi}{2}\frac{V(t)}{V_\pi}, \qquad (9)$$

leading in Figure 6 to an overall output phase delay of

$$\Delta\varphi = \Delta\varphi_1 - \Delta\varphi_2 = \pi\frac{V(t)}{V_\pi}, \qquad (10)$$

which conforms with the configuration of Figure 3 resp. Eq. (6).

## 3.2 Acousto-optic Modulator (AOM)

The intensity of a laser beam may also be controlled using an acousto-optic modulator (AOM). There, the light propagates through a transparent crystal or non-crystalline glass with an attached piezoelectric transducer which is in turn excited by an RF signal. Pressure fluctuations within the optical material cause a grating of its refractive index via the photo elastic effect. Thus, acoustic waves that travel through the material may deflect light, which is known as the acousto-optic effect. Here we only consider bulk wave acousto-optic devices, which are the most common, while also surface acoustic wave and optical guided devices exist [11], [7]. While this review is focused on modulating the power of a laser beam, AOMs are also used as optical switches, isolators, and scanners amongst others [1].

The change of the refractive index by an acoustic beam with power $P_a$, length $l$ and height $h$ can be described as

$$\Delta n = \sqrt{\frac{1}{2}M_2\frac{P_a}{lh}} \qquad (11)$$

Fraunhofer IIS 2024    Bias Control and Linearization of the Transfer Function of    7 | 22
Electro-optic and Acousto-optic Modulators

using the material's acousto-optic figure of merit

$$M_2 = \frac{n^6 p^2}{\varrho v_a^3} \tag{12}$$

that depends on the refractive index of the unstrained medium $n$, the appropriate element of the photo elastic tensor $p$, the mass density $\varrho$ and the acoustic velocity $v_a$ [2]. Although this change $\Delta n$ is in the order of $10^{-4}$ for an acoustic power density of $100 W/cm^2$, the overall optical effect can be significant when these minor changes accumulate.

There are two main types of diffraction that may be observed, namely Raman-Nath diffraction and Bragg diffraction, as shown in Figure 7. Additionally, an AOM may be operated in the near-Bragg regime, which denotes the transition of both configurations, where only a few orders of light are significant. For an overview of additional AOM types it is referred to [2].

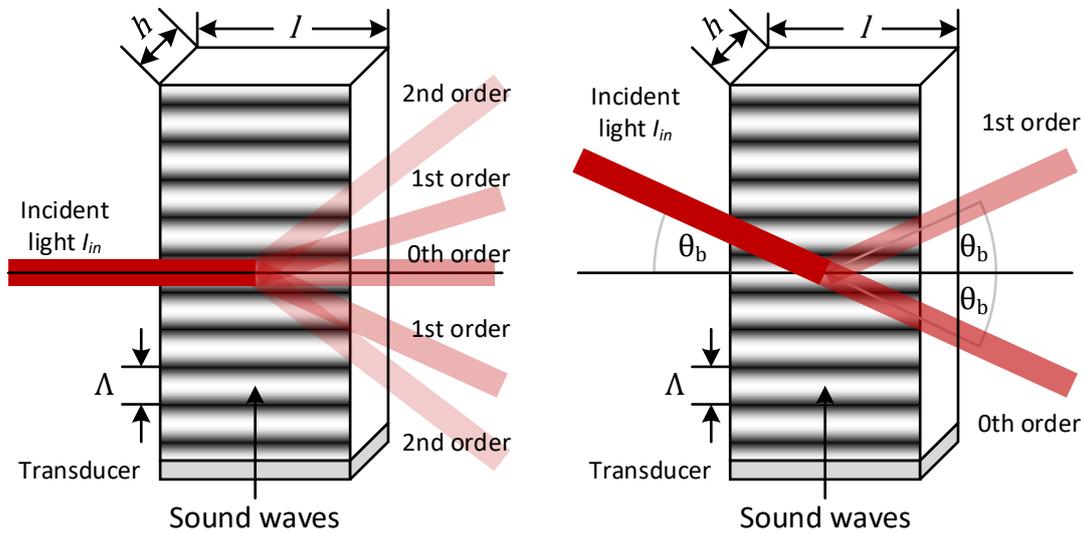

*Figure 7: Acousto-optic modulator in the Raman-Nath configuration (left) and Bragg configuration (right).*

The acousto-optic modulator operates in the Raman-Nath configuration, if the modulator's interaction length

$$l \ll \frac{\Lambda^2}{\lambda} \tag{13}$$

is very short compared to the acoustic wavelength $\Lambda$, normalized by the optical wavelength $\lambda$ within the modulator material. The light is then diffracted into different orders $m$ at angles conforming with

$$\sin \theta_m = \frac{m \lambda_0}{\Lambda}, \ m = 0, \pm 1, \pm 2, \dots \tag{14}$$

and of distinct intensities. Commonly the 0-th order beam is used as the output and its intensity is given by

$$I_0/I_{\text{in}} = |J_0(\Delta \phi')|^2 \tag{15}$$

with the ordinary Bessel function of the 0-th order $J_0$ and the unmodulated optical beam intensity $I_\text{in}$. Due to their comparatively low maximum intensity of about 34%, AOMs are usually not operated in the Raman-Nath regime [11].

Bragg-type modulators on the other hand depend on multiple diffractions of the optical beam at the acoustic wave and thus a relatively long interaction length

$$l \gg \frac{\Lambda^2}{\lambda} \quad (16)$$

is required. Here, the optical beam should incident at the Bragg angle $\theta_b$, conforming with

$$\sin \theta_b = \frac{\lambda}{2\Lambda} = \frac{\lambda_0}{2n\Lambda}, \quad (17)$$

where $\lambda_0 = \lambda n$ denotes the light's wavelength in vacuum. Thus, the incremental reflections at the material's reflective index grating result in a $2\pi$ phase shift which leads to constructive interference and only a single diffracted output. The diffracted light's intensity is strongly dependent on this angle and a deviation from this angle by $|\theta - \theta_b| \approx \lambda/(2l)$ already leads to a vanishing signal [1]. While the acoustic wave usually travels relative to the optical beam, this effect is negligible for most applications due to their long wavelength compared to the optical period. However, if the incident light beam points towards the sound wave vector, as illustrated in Figure 9, the diffracted light's frequency is upshifted by the acoustic frequency. Otherwise, when the incident light makes an acute angle with the sound wave, the diffracted light's frequency is downshifted by the acoustic frequency. Due to its lower optical losses, Bragg diffraction is more favorable and commonly used although its acousto-optic requirements limit the useable acoustic frequency range.

For plane sinusoidal acoustic and optical waves, the diffracted light's normalized intensity is given by

$$\frac{I_1}{I_{in}} = \eta \operatorname{sinc}^2 \sqrt{\eta + \left(\frac{\Delta k l}{2}\right)^2} \quad (18)$$

with $\operatorname{sinc}(x) = \sin(x)/x$, the parameter

$$\eta = \frac{\pi^2}{2\lambda_0^2} M_2 P_a \frac{l}{h} \quad (19)$$

and the momentum mismatch $\Delta k = |\boldsymbol{k_i} + \boldsymbol{k_a} - \boldsymbol{k_d}|$ [1]. It denotes the magnitude of the momentum mismatch among the incident light $\boldsymbol{k_i}$, the acoustic wave $\boldsymbol{k_a}$ and the diffracted light $\boldsymbol{k_d}$. If the Bragg condition is satisfied, $\Delta k = 0$ holds and Eq. (18) simplifies to

$$\frac{I_1}{I_{in}} = sin^2 \sqrt{\eta}, \quad (20)$$

which may be approximated linearly by $I_1/I_{in} \approx \eta$ for small values of $\eta \ll 1$ and is termed weak interaction.

If the momentums are not perfectly matched, we may approximate the first order's light intensity in Eq. (18) by

$$\frac{I_1}{I_{in}} \approx \sin^2 \sqrt{\eta} \operatorname{sinc}^2 \frac{\Delta k l}{2} = \sin^2 \sqrt{\eta} \operatorname{sinc}^2 \pi \psi \quad (21)$$

using the normalized phase mismatch for the general case

$$\psi = \frac{\Delta k l}{2\pi} = \frac{l}{2\lambda_0 n_d}\left(\left(\frac{\lambda_0}{\Lambda}\right)^2 - 2n_i\frac{\lambda_0}{\Lambda}\sin\theta_i + n_i^2 - n_d^2\right) \qquad (22)$$

at incident angle $\theta_i$ and where $n_i$ and $n_d$ denote the refractive indices for the incident and diffracted light respectively [11]. For isotropic materials these refractive indices $n = n_i = n_d$ coincide, while they differ for anisotropic materials. In this case, two different wave vectors lead to conservation of momentum and thus efficient diffraction, which enables devices of increased modulation bandwidth [7]. The phase mismatch of Eq. (22) for a specific incident angle $\theta_i$, corresponding to an acoustic center frequency, can be inserted into Eq. (21) to estimate the diffraction efficiency for adjacent spectral components. Under the assumption of weak interaction and thus $\eta \ll 1$, the behavior for finite geometries may therefore be traced analytically in terms of the Fourier analysis as outlined in [11].

Amplitude modulated acoustic signals generally have spectra of non-zero bandwidth, while the Bragg condition is only satisfied for its corresponding frequency in isotropic materials. Thus, an incident optical plane wave ideally only interacts with this frequency component of the acoustic wave, leading to a zero bandwidth and monochromatic diffracted light [1].

In practice, the idealized assumption of planar optical and acoustic waves is usually not met. However, we may decompose an incident gaussian laser beam of width $d$ into planar waves with an angular divergence

$$\delta\theta_0 = \frac{4\lambda_0}{\pi n d}. \qquad (23)$$

When interacting with a single frequency planar acoustic wave, only the optical planar wave satisfying the Bragg condition gets reflected, resulting in planar output waves. Thus for modulated acoustic signals, diffraction occurs for a range of acoustic frequencies, which leads to a diffracted light beam [1]. As the acoustic wave, generated from a flat transducer of length $l$, is assumed to have a beam divergence

$$\delta\theta_a = \Lambda/l \qquad (24)$$

in the interaction plane, the Bragg condition is met for even more spectral components [11]. However, there is a tradeoff between the modulation bandwidth and the deflection efficiency. A corresponding focused-beam-type AOM is sketched in Figure 8.

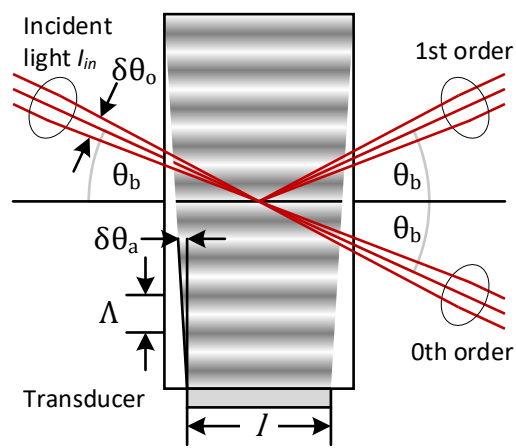

*Figure 8: Focused-beam-type AOM used for upshifted diffraction [13].*

The characteristics of an AOM mainly depends on two parameters, which should be adapted to meet the desired specifications. For a start, the acousto-optical interaction is characterized by the beam divergence ratio

$$a = \frac{\delta\theta_0}{\delta\theta_a} = \frac{4\lambda_0 l}{\pi n d \Lambda}, \quad (25)$$

which should be close to unity for a modulator configuration [11]. Furthermore, the switching time of a diffracted laser beam's intensity is limited by the acoustic transit time

$$\tau_a = d/v_a \quad (26)$$

which denotes the time that the acoustic wavefront, traveling at speed $v_a$, requires to pass the laser beam of width $d$. Additionally, the piezoelectric transducer's response time and the traveling time of the acoustic wave to the beam position also contribute to the system's overall response time [2].

The diffracted light's intensity may be modulated by means of amplitude modulation. The AOM's modulation bandwidth is characterized by the 3dB modulation frequency $f_m^{3dB}$. It denotes the spectral distance from the RF center frequency $f_0 = v_a/\Lambda$ at which the diffracted beam's intensity is reduced by 50% and may be approximated by

$$f_m^{3dB} \approx \begin{cases} 0.75/\tau_a, & a \ll 1 \\ 0.7/\tau_a, & 0.67 < a < 1.5 \end{cases} \quad (27)$$

depending on the divergence ratio $a$ and rise time $\tau_a$ [11]. For Gaussian input beams, the frequency response is also stated in [12].

By decreasing the beam width, the switching time increases which in turn increases the modulation bandwidth. However, this also increases the optical spread and thus the divergence ratio and the high peak intensity may damage the optical material. In order to limit the eccentricity to the diffracted beam to 10%, the divergence ratio should be 0.67. The divergence ratio $a = 1.5$ maximizes the modulation bandwidth and the rise time $t_r = 0.85$ from 10% to 90% of the steady state diffraction intensity [11].

Furthermore, in order to be able to separate the diffracted beam from the incident beam, the Bragg angle should exceed optical beam divergence. Thus, the acoustic center frequency should comply with

$$f_0 = \frac{v_a}{\Lambda} > \frac{8}{\pi\tau} \quad (28)$$

and therefore be at least four times the maximum modulation frequency [11]. In order to obtain higher bandwidths, phased array transducers or a series of multiple tilted AOMs may be used [2].

The transducer consists of a top and bottom electrode layer to which an RF signal is applied and with a piezoelectric layer in between. While one may consider the transducer as a simple converter from electrical energy to acoustic energy with a certain efficiency, a frequency dependent two-port description of the electrical and mechanical network is more adequate at high frequencies [12]. To operate the AOM efficiently, the system's impedance should be matched to the RF driver. For this a matching network is used, which may consist of lumped elements although microstrip techniques may be more suitable at high frequencies [12]. Additionally, the optical material's acoustic attenuation will affect the system's performance [11] and the acousto-optic medium will become nonlinear for high acoustic power densities [7].

For progressive wave AOMs, the top part is cut at an angle and serves as a termination by absorbing parts of the acoustic wave and reflecting the acoustic wave away from the incoming light. The top of a standing wave AOMs on the other hand is parallel to the piezoelectric transducer in a distance of an integer multiple of the acoustic wavelength. Thus, while standing wave AOMs are very efficient, they only function for acoustic frequencies satisfying these conditions and are quite sensitive to temperature variations. The standing wave leads to two doppler shifted spectral components in the diffracted beam.

## 3.3 Conclusions regarding Control Voltage Generation for EOM and AOM

According to the physical effects described above, both EOM and AOM have basic nonlinear transfer functions from control voltage input to optical output power (assuming constant optical input power). In addition, linear distortions can be expected for the EOM as an electric field has to be switched in a structure that can be considered as a capacitance. Linear distortions also occur in the AOM due to the acoustic transit time and the frequency dependency of the piezoelectric transducer. In summary, this results in a nonlinear system with memory.

If a precise power envelope is targeted at the output of an optical modulator, the voltage at its control input must be designed accordingly. From an application point of view, this can be achieved with a linearization system that models the transfer function of the optical modulator in its linear and nonlinear distortions and generates a correspondingly predistorted control voltage based on the desired output waveform. Regarding the EOM, the drift of the bias voltage must also be taken into account and compensated for, which is performed by a separate control loop.

# 4 Pulse Linearization of Modulator Control Voltage to Optical Output Power

## 4.1 Structure of the Modulator Control System applied for Pulse Linearization

In this section the basic structure and functionality of the proposed modulator control system are described.

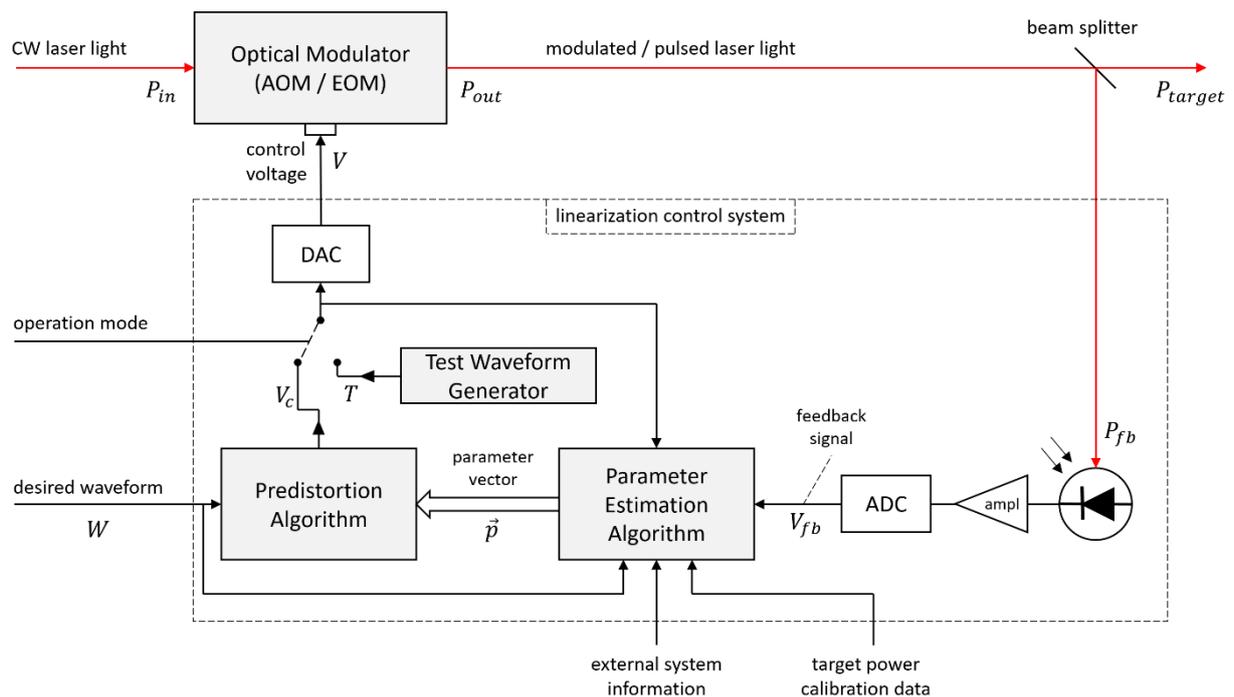

Figure 9: Schematic diagram of the proposed linearization control system

Figure 9 depicts the schematic diagram of the proposed modulator control system. At this level, no distinction is made whether an AOM or EOM is used as modulator. It is the main task of the above linearization system that both amplitude and envelope of the target output power $P_{target}$ correspond to an arbitrary desired waveform $W$ as accurately as possible. This is achieved by two main functional elements: a parameter-controlled predistortion algorithm and a parameter adaptation algorithm, which repeatedly adjusts the predistortion parameters based on evaluation of a feedback signal.

The continuous-wave (CW) laser light with input power $P_{in}$ passes through the modulator, which is controlled by the control voltage $V$. It is not required that $P_{in}$ is strictly constant over a long time. However, it is assumed that the laser light power is not modulated itself and power variations are relatively slow, i.e. they do not occur within the duration of an active optical modulation process.

$P_{out} = P_{out}(t)$ denotes the laser light power at the output of the optical modulator. It may typically be shaped as a pulse or a sequence of pulses (pulse train). A beam splitter splits the modulated laser beam into a component destined for the target and a component fed into a detector (e.g. photo diode), which converts a part of the modulator output power into an electrical signal. It is assumed that the splitting ratio $R_{split} = P_{fb}/P_{target}$ is constant and known (e.g. by calibration measurements). After suitable amplification, low pass filtering (not depicted) and A/D-conversion (ADC) a discrete-time feedback signal $V_{fb} = V_{fb}[k]$ is available, where $k$ is the sample index.

Ideally, the feedback signal $V_{fb}$ is strictly proportional to the target output power, i.e. $V_{fb}[k] = K \cdot P_{target}(t = k \cdot T_s)$, where $K$ is a known constant and $T_s$ is the sampling period of the ADC. Although this may not be fully achieved in a real-world system, high linearity, low noise and low amount of linear distortion of the overall feedback path (beam splitter to ADC output) are crucial for the linearization performance.

The core of the linearization system is the predistortion algorithm. Its task is to generate a modulator control voltage in such a way that the target power $P_{target}$ (used as input for quantum operations) corresponds to a desired waveform $W$, both in terms of shape and amplitude. This implicitly includes that e.g. the energy of modulated light pulses can be controlled. Regarding the "electro-optic" transfer function from control voltage input $V$ to optical output power $P_{out}$ the optical modulator introduces distortion as shown in section 3. The predistortion algorithm can thus be considered to implement the inverse electro-optic transfer function in order to linearize the transfer function from input waveform $W$ to target output $P_{target}$. In the proposed implementation the predistortion algorithm consists of a given fixed structure and a set of variable parameters represented by a parameter vector $\vec{p}$ (Figure 9).

The parameter vector is provided to the predistortion algorithm by the adaptation algorithm, which evaluates the feedback signal $V_{fb}$ (ideally representing the actual modulator output) in comparison to the desired waveform $W$. Due to a variety of physical effects (change of internal and environmental conditions, such as e.g. thermal drift etc.) the properties of the optical modulator are time variant. Therefore, the parameter vector $\vec{p}$ has to be updated either periodically, event driven or on demand. In active quantum computing operation mode this is based on the evaluation of input waveforms $W$. In addition, a test waveform generator may provide a test waveform $T$, which is applied exclusively for the purpose of updating the parameter vector $\vec{p}$. In that case, it must be ensured that this does not affect the subsequent part of the quantum computer.

## 4.2 Overview of Predistortion Algorithm Candidates and Parameter Estimation Strategies

In this section an overview of potential predistortion algorithms is given which may be applied for the linearization structure described in section 4.1. As explained above, the predistortion ideally implements the inverse transfer function of the optical modulator, so that in a first step a suitable model must be selected for the modulator transfer function, which is then inverted for linearization.

### 4.2.1 Overview on common nonlinear Memory Models

In section 3.1 and 0 it was shown that the transfer functions of the considered optical modulators (AOM and EOM) include both nonlinear contributions and memory effects. The latter means, that the output $P_{out}(t = t_0)$ not only depends on the instantaneous input voltage $V(t_0)$, but also on the input voltage $V(t < t_0)$. In a real world system it can be assumed that the practically relevant memory length is restricted to a time period $T_{mem}$, i.e. $P_{out}(t_0) = f(V(t))$ with $t \in [t_0 - T_{mem}; t_0]$, where $f(.)$ is a nonlinear function.

In the following we consider a discrete-time system model based on the assumption that input and output signal of the real-world system are sampled (A/D converted) and generated (D/A converted) at sufficiently high sampling frequency $f_s$, specifically w.r.t. fulfilling the sampling theorem. Accordingly, the system memory of length $T_{mem}$ is represented by a discrete-time $M$-tap memory. For the sake of better readability, input and output of the nonlinear system models described below are denoted by dimensionless variables $x$ and $y$ instead of their physical equivalents $V$ and $P_{out}$.

Generic Volterra series model
The most general form for a model implementing nonlinearity with finite memory $M$ is described by the Volterra series, following a similar approach as the well-known Taylor series. According to [16,17,18] the output $y(n)$ of a causal, stable, time-invariant, finite memory, discrete-time system can be written as

$$y(n) = \sum_{k=1}^{K} y_k(n), \quad (29)$$

where

$$y_k(n) = \sum_{m_1=0}^{M-1} \cdots \sum_{m_k=0}^{M-1} h_k(m_1, \cdots m_k) \prod_{l=1}^{k} x(n - m_l) \quad (30)$$

is the $k$-dimensional convolution of the input sequence $x(n)$ with the $k^{th}$ order Volterra kernel $h_k$.

Although (30) can be rewritten in an alternative form applying only one-dimensional convolutions one of the problems inherent to the Volterra representation is the computational complexity involved in calculating the output due to the large number of parameters in the Volterra kernels [16,17]. Therefore, some simplified structures can be used as approximation, depending on the required model accuracy and on the characteristics of the system to be modeled (see Figure 10).

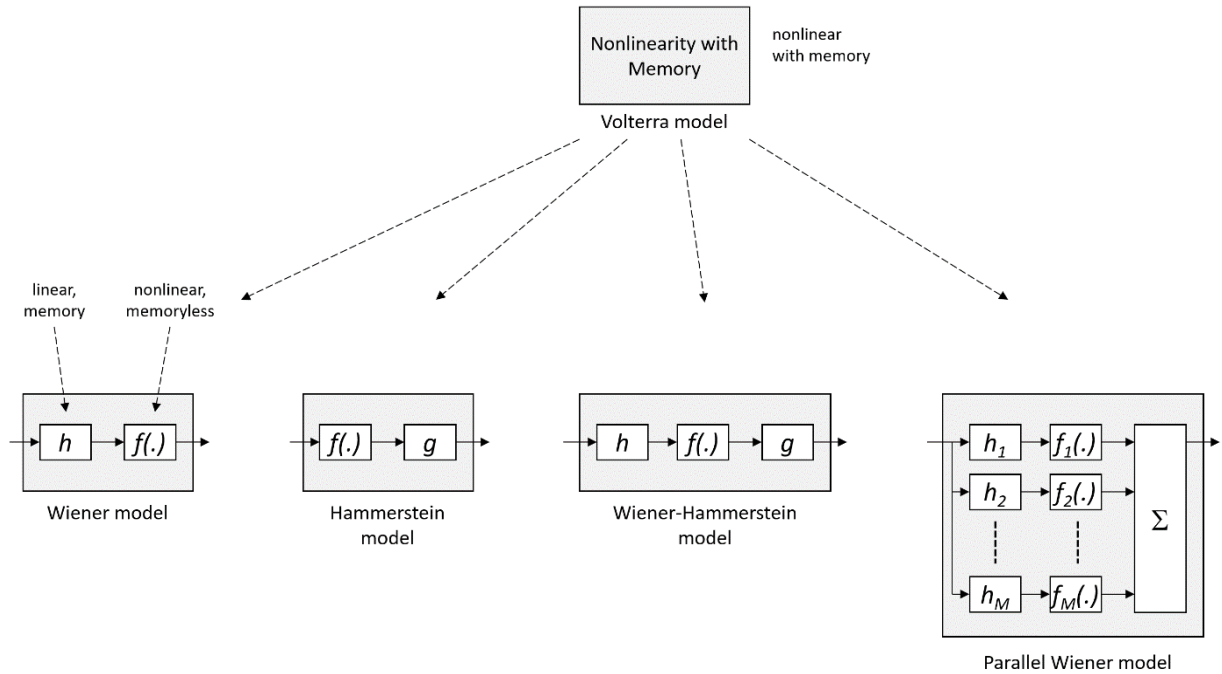

*Figure 10: Selection of some common nonlinear memory models*

Wiener model
The Wiener model is one of the simplest structures to combine memory effects with nonlinearity. It consists of a linear filter $h$ followed by a memoryless nonlinearity $f(\cdot)$, see Figure 10 (2$^{nd}$ from left). The output can be written

$$y(n) = \sum_{k=1}^{K} a_k \left( \sum_{m=0}^{M-1} h(m) x(n-m) \right)^k . \tag{31}$$

The number of parameters ($K + M$) is much smaller than in the generic Volterra model of same memory length. However, the Wiener model depends nonlinearly on the coefficients $h(m)$, making their estimation more troublesome than for models that are linear in their parameters [17].

Hammerstein model
In the comparably simple Hammerstein model, the order of memory and nonlinearity is reversed compared to the Wiener model (see Figure 10). The output is given by

$$y(n) = \sum_{m=0}^{M-1} g(m) \sum_{k=1}^{K} a_k x^k(n-m). \tag{32}$$

In contrast to the Wiener model, the Hammerstein model is linear in all parameters ($g(m), a_k$). Under certain assumptions Hammerstein and Wiener models can form mutual inverses.

Wiener-Hammerstein model
The Wiener-Hammerstein (or Hammerstein-Wiener) model has the concatenated structure of a linear filter, a memoryless nonlinearity, and another linear filter (see Figure 10, 2$^{nd}$ from right). Its output is given by

$$y(n) = \sum_{m_2=0}^{M-1} g(m_2) \sum_{k=1}^{K} a_k \left( \sum_{m_1=0}^{M-1} h(m_1) x(n - m_1 - m_2) \right)^k . \qquad (33)$$

This model has the undesired property of being nonlinear in the parameters $h(m_1)$ same as the Wiener model. The Wiener-Hammerstein model essentially has the similar or same structure as neural networks or multilayer perceptrons. This similarity enables one to benefit from both the behavioral model in system identification and powerful tools in machine learning [19].

Beside the above introduced common simplifications of the generic Volterra model there are further structures like e.g. the parallel Wiener model (Figure 10 bottom right) or a structure denoted as memory polynomial [19], some of which can be considered as generalizations of a Wiener or Hammerstein model.

### 4.2.2  Model Selection Criteria

The selection criteria for a suitable nonlinear model are primarily modeling accuracy and implementation complexity. Further important aspects are stability, flexibility (e.g. regarding hardware changes), implementation costs.

Modeling accuracy
The accuracy of the nonlinear modulator model is the basis for the appropriate operation and precision of the predistortion. Therefore, the modeling accuracy is crucial for the accuracy of the complete linearization process, i.e. for the degree to which the target laser light power $P_{target}$ matches the desired waveform $W$. In ideal case (ignoring processing delays), this would result in $P_{target}(t = n \cdot T_s) = C \cdot W(n)$, where $n$ is an integer time index, $T_s$ represents the sampling interval of $W$, and $C$ denotes a known conversion constant between the unitless waveform $W$ and the physical light power $P_{target}$.

Depending on the requirements of the quantum computing system processing the modulated laser light a suitable error metric $E_\Delta = f_\Delta(P_{target}, C \cdot W)$ has to be defined, which may e.g. be a least mean squares metric.

Implementation complexity
Implementation complexity refers to both complexity of the predistortion algorithm as well as complexity of the related parameter estimation algorithm.

Regarding predistortion, two different modes can be considered: stream processing or block-wise processing. The former means that a continuous input stream $W$ has to be predistorted permanently in real-time, requiring the corresponding processing resources. In block-wise processing, $W$ may hold a single pulse or a pulse train of limited length, followed by an inactive time period with $P_{target} = 0$. Assuming that $W$ is known sufficiently earlier w.r.t. the desired modulation time, the predistortion algorithm may run "offline", i.e. slower than real-time. This mode requires both sufficient lead time for the predistortion and sufficiently long inactive time periods ($P_{target} = 0$). It allows for either a higher computational complexity of the predistortion algorithm or for lower processing resources.

Due to changes in environmental conditions, device temperature, aging etc. the parameters of the nonlinear model have to be updated regularly by the parameter estimation algorithm. Depending on the nonlinear model and the applied estimation algorithm the parameter estimation is expected to be of considerable computational complexity. Therefore, it may be performed in non-real-time mode. A set of updated parameters is provided to the predistortion as soon as its computation is completed. There has to be a tradeoff between parameter estimation frequency (update rate) and estimation complexity. In order to save computational load it may be an option to update only a subset of parameters at a time.

### 4.2.3 Basic Approaches for Predistortion Function and Parameter Estimation

As stated above, the predistortion ideally is the inverse of the nonlinear transfer function of the optical modulator. The conventional approach is to first model the nonlinear transfer function (e.g. using one of the above models) and, based on that, calculate the inverse of the nonlinear model for use as a predistortion function. For Volterra models, there is a technique known as $p^{th}$-order inverse [18], but it is of significant computational complexity.

An alternative approach is proposed in [20] which avoids computing the inverse of a nonlinear model. Its basic structure is depicted in Figure 11.

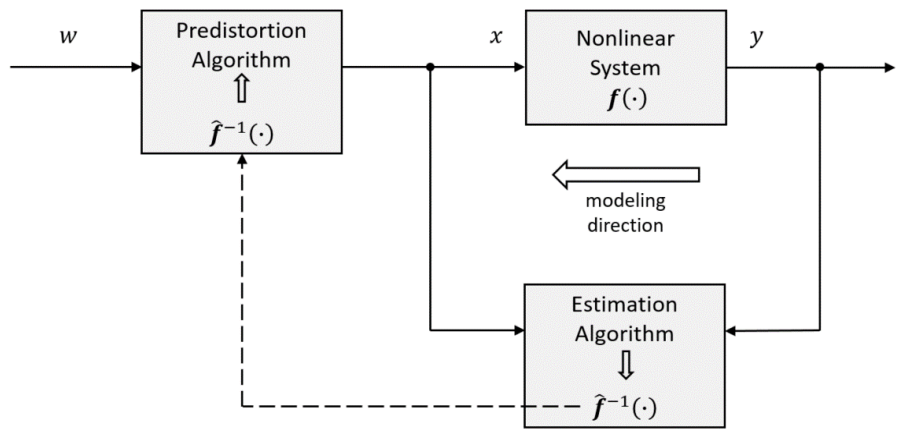

*Figure 11: Predistortion applying inverse nonlinear modeling*

The fundamental idea is to model the nonlinear system with transfer function $y = f(x)$ in inverse direction. Based on the nonlinear system input $x$ and the observed output $y$ the estimation algorithm models the inverse transfer function $x = f^{-1}(y)$, resulting in the estimated inverse function $\hat{f}^{-1}(\cdot)$. In [17] this is denoted as "post distortion". Any of the above presented models can be used for the inverse model. The identical model (structure and estimated parameters) is then applied as predistortion function, i.e. the predistortion function is written as $x = \hat{f}^{-1}(w)$. In [18] it is shown that the $p^{th}$-order post-inverse of a general Volterra system is identical to its $p^{th}$-order pre-inverse.

The inverse modeling approach has been successfully applied to the predistortion of power amplifiers, e.g. in wireless communication systems. If the structure of the inverse model is a linear weighted combining of nonlinear signals (being linear in its coefficients) the estimation algorithm can be implemented e.g. as a least-squares solution at comparably moderate complexity.

## 4.3 Aspects of Predistortion Parameter Updates and Test Waveform Generation

With regard to predistortion parameter updates, both the frequency of the updates and the underlying waveforms have to be considered. Parameter update in this context denotes the execution of a parameter estimation and the corresponding parameter update in the predistortion unit.

The parameter update frequency depends on the time variance of the modulator characteristics (e.g. speed of thermal drift), the predistortion accuracy requirements, and the available computational resources. Both the complete set of parameters as well as a suitable subset of parameters may be estimated in an update process. Three update approaches can be distinguished:

- periodic updates,
- error metric based updates,
- event triggered updates.

While periodic updates are performed at fixed time intervals, error metric based updates are only initiated when the error metric, i.e. the deviation of the measured output waveform from the desired waveform, exceeds a certain level. In addition, parameter updates may also be triggered by system events or manually.

Parameter estimation may be performed based on either quantum computing waveforms or on specific estimation waveforms. The former are designed for active quantum computing operations and thus may not be ideally suited for parameter estimation. In contrast, estimation waveforms are optimized specifically for that purpose. They can be applied only at times when quantum computing is not active and it has to be ensured (e.g. by an optical switch) that the quantum computer is not affected by laser light from the modulator output. Design criteria for estimation waveforms shall be such that fast and accurate modeling of the modulator transfer function is possible.

# 5  Compensation of long-term Physical Parameter Fluctuations

For precise control of optical modulators, in addition to their basic nonlinear characteristics, other physical effects causing time-variant behavior must be taken into account. This specifically concerns the bias drift of electro-optic modulators and temperature dependencies of the devices in general. The adaptive predistortion proposed above includes a parameter update procedure, allowing to compensate for changes in this transfer function. However, a re-estimation of predistortion parameters may cause a considerable computational effort, especially for a multi-channel system. Therefore, well-known effects such as the EOM's bias drift can be handled in separate control loops of lower complexity.

Bias Drift (EOM)
As described in Eq. (8) in section 3.1, the phenomenon of bias drift $\varphi_{drift}$, caused for example by thermal effects, causes the transfer function of the MZM to drift over time and change the 4 optimum bias point settings like negative or positive slope quadrature, minimum and maximum point. The thermal bias drift phenomenon is mainly caused by the charge migration and the build-up of pyroelectric charges in the substrate due to environmental perturbations [1, 2, 4]. In the past, many attempts have been made to introduce techniques to minimize the bias drift through the introduction of new device fabrication, special design, and packaging [4, 9]. Even though these approaches improve the stability of the bias drift, additional precise bias control is still required in an MZM.

In general, there are mainly two categories of bias control technologies, i.e., the optical power-based technique [4,10] and the pilot-based technique [4,10,22,23]. The basic principle of the optical power-based technique is to monitor the power of the input or output optical signal of the MZM or the power ratio between them. The variation of the power or the power ratio indicates the drifting of the bias point. By feeding the drift information back to the MZM, the bias point of the modulator can be locked. Further improvements can be reached by using a microcontroller that can measure and update the MZMs transfer function and the PDs output voltage frequently.

The techniques in the second category make use of the nonlinearity of the MZM to compensate for the bias drift. When a small pilot RF tone is added to bias voltage, some high order harmonics (second- and/or third-order) may be generated due to MZM's nonlinearity of its transfer function. The RF power of the high order harmonic is then used to generate an error signal, which is converted to a control signal to compensate for bias drift. Preferably, the 4 above named operation points of the MZM will be used for this purpose [22,23].

Temperature Drift (AOM)
In the long run, the optical and acoustic power absorbed by the AOM will lead to a temperature increase. As the material's speed of sound is temperature dependent and a temperature gradient also impairs the refractive index, this affects its acousto-optic figure of merit [7] [21]. Thus, the deflection angle, diffraction efficiency and the optical beam may be impaired [7]. We may combat this temperature drift by modeling the temperature dependence, appropriate cooling solutions or regular recalibration of the control loop. By feeding a constant acoustic power to the device, the temperature variation may be reduced. The output signal on the target may still be extinguished, when tuning the RF frequency to a different Bragg angle at which the diffracted beam is blocked. One additional approach is to estimate several predistortion parameter sets, each valid for a different device temperature, and select the best-fitting set as a function of the actual device temperature. This involves an increased initial effort, but the frequency of parameter estimation during operation can be reduced if temperature is the main influencing factor behind the time-varying behavior of the modulator.

The above brief review describes different basic approaches to generate AOM and EOM control signals in such a way that almost arbitrary target output waveforms are achieved with high accuracy.

## Acknowledgements

All authors acknowledge funding by the BMBF project MUNIQC-ATOMS (Grant No. 13N16071).

## References

[1]  B. E. A. Saleh, and M. C. Teich, "Fundamentals of Photonics", 3rd ed. New York, NY: John Wiley and Sons, 2019
[2]  R. G. Hunsperger, "Integrated optics: Theory and technology", 6th ed. New York: Springer, 2009
[3]  A. Chen and E. J. Murphy, "Broadband Optical Modulators – Science, Technology, and Applications", CRC Press, 2012
[4]  Y. Fu, X. Zhang, B. Hraimel, T. Liu, and D. Shen, "Mach-Zehnder: A review of Bias Control Techniques for Mach-Zehnder Modulators Photonic Analog Links", IEEE Microwave Magazine, vol. 14, no. 7, pp. 102-107, 2013

[5] H. K. Shankarananda, S. S. Shreyas, and B. Guruprasad, "External Modulators and Mathematical Modeling of Mach-Zehnder Modulator", International Journal of Innovative Science, Engineering & Technology, vol. 3, issue 12, pp. 214-220, Dec 2016

[6] C. H. Park, M-K. Woo, P. K. Park, S. W. Jeon, H. Jung, S. Kim, and S. W. Han, "Experimental Demonstration of an Efficient Mach-Zehnder Modulator Bias Control for Quantum Key Distribution Systems", Electronics, 11, 2207, 2022, doi: 10.3390/electronics11142207

[7] N. J. Berg, and J. M. Pellegrino, "Acousto-optic signal processing: Theory and Implementation", 2nd ed. New York, Marcel Dekker Inc., 1996

[8] K. Iizuka, "Engineering Optics", 4th ed. New York: Springer, 2007

[9] X. Huang, Y. Liu, D. Tu, Z. Yu, Q. Wei, and Z. Li, "Linearity-Enhanced Dual-Parallel Mach-Zehnder Modulators based on a Thin-Film Lithium Niobate Platform", Photonics, 2022

[10] Z. Pan, S. Liu, N. Zhu, P.Li, M. Liu, L. Yang, C. Du, Y. Zhang, and S. Pan, "Arbitrary Bias Point Control for Mach-Zehnder Modulator using a Linear-Frequency Modulated Signal", IEEE Photonics Technology Letters, Vol. 33, No. 11, June 2021

[11] M. Bass and Optical Society of America, Eds., "Handbook of optics", 2nd ed. New York: McGraw Hill, 1995

[12] Jr. Eddie, H. Young, and S.-K. Yao, "Design considerations for acousto-optic devices", Proc. IEEE, vol. 69, no. 1, pp. 54–64, Jans. 1981, doi: 10.1109/PROC.1981.11920

[13] I. C. Chang, "I. Acoustooptic Devices and Applications", IEEE Trans. Sonics Ultrason., vol. 23, no. 1, pp. 2–21, Jan. 1976, doi: 10.1109/T-SU.1976.30835

[14] F. Griesmer, "Optimizing Mach-Zehnder Modulator Designs with COMSOL Software", Comsol Blog, April 2014

[15] J. Porins, V. Bobrovs, "Realization of HDWDM Transmission System with the Minimum Allowable Channel Interval", chapter 9 in book "Optical Communications Systems", March 2012, doi: 10.5772/28784

[16] G. M. Raz and B. D. Van Veen, "Baseband Volterra filters for implementing carrier based nonlinearities", IEEE Trans. Signal Processing, vol. 46, pp. 103-113, Jan. 1998

[17] Dennis R. Morgan, Zhengxiang Ma, Jaehyeong Kim, Michael G. Zierdt, and John Pastalan, "A Generalized Memory Polynomial Model for Digital Predistortion of RF Power Amplifiers", IEEE Trans. Signal Processing, vol 54, pp 3852-3860, 2006

[18] M. Schetzen, "The Volterra and Wiener Theories of Nonlinear Systems", New York: Wiley, 1980

[19] Takeo Sasai, Masanori Nakamura, Etsushi YamazakiI, Asuka Matsushita, Seiji Okamoto, Kengo Horikoshi, and Yoshiaki Kisaka, "Wiener-Hammerstein model and its learning for nonlinear digital pre-distortion of optical transmitters", Optics Express Vol. 28, Issue 21, pp. 30952-30963, 2020

[20] C. Eun and E. J. Powers, "A new Volterra predistorter based on the indirect learning architecture", IEEE Trans. Signal Processing, vol. 45, pp. 223-227, Jan. 1997

[21] X. Zhang, Y. Chen, J. Fang, T. Wang, J. Li, and L. Luo, "Beam pointing stabilization of an acousto-optic modulator with thermal control," Opt. Express, vol. 27, no. 8, p. 11503, Apr. 2019, doi: 10.1364/OE.27.011503

[22] Y. Li, Y. Zhang and Y. Huang, "Any bias point control technique for Mach-Zehnder modulator", IEEE Photonics Technology Letters, Vol. 25, No. 24, Dec. 2013

[23] L. L. Wang and T. Kowalcyzk, "A versatile bias control technique for any-point locking in lithium niobate Mach-Zehnder modulators", J. Lightwave Technol., Vol.28, No. 11, P. 1703–1706, Jun. 2010